# Estimation of intrinsic diffusion coefficients in a pseudo-binary diffusion couple

Sangeeta Santra and Aloke Paul*

Department of Materials Engineering, Indian Institute of Science, Bangalore, India

*Corresponding author: E-mail: aloke@materials.iisc.ernet.in, Tel.: 918022933242, Fax: 91802360 0472

**Abstract**

Major drawback of studying diffusion in multi-component systems is the lack of suitable techniques to estimate the diffusion parameters. In this study, a generalized treatment to determine the intrinsic diffusion coefficients in multi-component systems is developed utilizing the concept of pseudo-binary approach. This is explained with the help of experimentally developed diffusion profiles in the Cu(Sn, Ga) and Cu(Sn, Si) solid solutions.

---

Estimation of the diffusion coefficients following the conventional approach in a binary system implementing the diffusion couple technique is rather straightforward. It is a stringent process in a ternary system and impossible in multi-component systems [1]. As an alternate approach, the concept of the average effective interdiffusion coefficient was introduced by Dayananda and Sohn [2]. This is a kind of average interdiffusion coefficient of the main and cross interdiffusion coefficients over a wide composition range. Much recently, Chen et al. [3] proposed a method to estimate these parameters from a single diffusion couple in a ternary system, which otherwise needs two composition profiles from different diffusion couples to intersect at a composition at which these can be estimated. In the mean time, a simplified method named as pseudo-binary approach is followed to estimate the variation of the interdiffusion coefficients with composition from a single diffusion couple. Recently, one of the authors' of this manuscript reformulated this approach and explained the methodology [4], which was not used ideally in previous manuscripts [5-7]. To follow this approach, the diffusion couple in ternary or multi-component system should be prepared such that only two components develop the diffusion profiles whereas the composition of others remains constant throughout the interdiffusion zone.

An additional advantage of following this approach is presented in this manuscript based on the experimental results. The estimation methodology to determine the intrinsic diffusion coefficients is proposed, which is very difficult to estimate in a ternary system and becomes impossible with the increasing number of components following the conventional method. For example, to estimate the intrinsic diffusion coefficients in a ternary system, the inert markers (Kirkendall marker plane) should be found at the point of composition at which the intersection of two different diffusion profiles occurs from two different diffusion couples, which is impossible to design experimentally unless found incidentally. As a matter of fact, this was indeed found incidentally only once in a Fe-Ni-Al system [8]. However, it becomes a daunting task with an increase in number of components and nearly inconceivable following experimental approach. A strategy to estimate the ratio of intrinsic diffusion coefficients in a multi-component system from a single diffusion couple was proposed by Paul [4] in line compounds, in which different components occupy certain sublattices. In this manuscript, a generalized approach is proposed which could be used even in solid solutions with no preference for the components occupying different crystal lattice positions. For this, Cu-Sn-Ga and Cu-Sn-Si ternary systems are considered. It should be noted here that the same approach can be followed in systems with higher number of components provided only two are allowed to exhibit diffusion profiles in the interdiffusion zone. The systems chosen in this study are important for the growth of A15 intermetallic superconductors by the bronze technique [9].

Cu(8 at.%Sn) was prepared in an induction melting furnace and Cu(8 at.%Ga) & Cu(8 at.%Si) alloys were prepared in button form in an arc melting furnace under Ar atmosphere. These buttons were melted three times after flipping every time for a better homogeneity. All the alloys were then homogenized at 775°C (1048 K) for 50 h. The compositional homogeneity was examined in an EPMA (Electron Microprobe Analyzer) randomly at various spots and the deviation was found to be within ±0.2 at.% from the average reported composition. The alloys were then EDM (electro discharge machining) cut into pieces of 1 mm thickness. Cu(8Sn)/Cu(8Ga) and Cu(8Sn)/Cu(8Si) diffusion couples were then prepared in special fixtures after a standard metallographic preparation. $ZrO_2$ inert particles (~1 μm) were used as inert Kirkendall markers to locate the Kirkendall plane at which the intrinsic diffusivities could be estimated. The diffusion couples were annealed at 700°C (± 5°C) for 25 h in a calibrated vacuum furnace maintained at ~$10^{-4}$ Pa. After the heat treatment, the diffusion couples were cross-sectioned and the composition profiles across the interdiffusion zones were measured in EPMA with pure elements as standards after the metallographic treatment. The Kirkendall marker plane position was determined by detecting the location of $ZrO_2$ using the X-ray peaks originated from Zr.

The composition profiles of the interdiffusion zone are shown in Figure 1 (micrographs are shown in supplementary file). The location of the Kirkendall marker planes are indicated by "K". In a binary A-B system, the interdiffusion coefficient following the Wagner's approach [10] is expressed as

$$\tilde{D}(Y_i^*) = \frac{V_M^*}{2t\left(\frac{dY_i^*}{dx}\right)}\left[(1-Y_i^*)\int_{x^{-\infty}}^{x^*}\frac{Y_i}{V_M}dx + Y_i^*\int_{x^*}^{x^{+\infty}}\frac{(1-Y_i)}{V_M}dx\right] \quad (1a)$$

where $Y_i = \frac{N_i - N_i^-}{N_i^+ - N_i^-}$ is the composition normalizing variable, $N_i$ is the mole fraction of the component *i*, and $N_i^-$ and $N_i^+$ are the end member compositions on the left and right hand side of the diffusion couple, respectively, and hence $N_i^- \leq N_i \leq N_i^+$ in this case. $V_M$ (metres$^3$/mol) is the molar volume, *t* (seconds) is the annealing time, *x* (metres) is the position parameter. The terms $x^{-\infty}$ and $x^{+\infty}$ are the position parameters at the unaffected parts of the diffusion couple (left and right hand side, respectively) such that $x^{-\infty} \leq x \leq x^{+\infty}$. The asterisk represents the position of interest.

Interdiffusion coefficients can be estimated at every composition in the interdiffusion zone. On the other hand, the ratio of the intrinsic diffusivities can be estimated only at the Kirkendall marker plane using the relation proposed by van Loo [11] expressed as

$$\frac{D_A}{D_B} = \frac{\overline{V}_A}{\overline{V}_B}\frac{\left[N_A^+\int_{x^{-\infty}}^{x_K}\frac{Y}{V_M}dx - N_A^-\int_{x_K}^{x^{+\infty}}\frac{1-Y}{V_M}dx\right]}{\left[-N_B^+\int_{x^{-\infty}}^{x_K}\frac{Y}{V_M}dx + N_B^-\int_{x_K}^{x^{+\infty}}\frac{1-Y}{V_M}dx\right]} \quad (1b)$$

Following the absolute values of the intrinsic diffusion coefficients can be estimated using the relation [12]

$$\tilde{D} = C_A \overline{V}_A D_B + C_B \overline{V}_B D_A \tag{1c}$$

where $C_i = N_i / V_M$ is the concentration and $\overline{V}_A$ and $\overline{V}_B$ are the partial molar volumes at the Kirkendall marker plane composition of components *A* and *B* respectively. These relations are developed using $N_A + N_B = 1$. In a pseudo-binary system, two components, like a binary system, develop a diffusion profile keeping composition of other components constant over the whole interdiffusion zone. In the systems considered in this study, the composition of Cu is constant and the other two components develop the diffusion profiles as shown in Figure 1. Therefore, the same relations shown above can be used to estimate the diffusion parameters; however, the composition profiles should be modified. For the estimation of the interdiffusion coefficients, the compositions should be normalized such that *Y* varies from 0 to 1. On the other hand, for the use of the intrinsic diffusion coefficients composition profiles of the diffusing components should be made such that the total of the fraction of compositions considered for the estimation is 1. Since, the systems are complete solid solution within the considered composition range without any preference of any component for a particular lattice; one can achieve this by adding equal amount of non-diffusing component to the diffusing components. Therefore, the modified compositions of A and B where C is non-diffusive can be expressed as -

$M_A = N_A + \frac{1}{2} N_C$ -----(2a) $M_B = N_B + \frac{1}{2} N_C$ ----- (2b) such that $M_A + M_B = 1$.

Following the modified normalized composition variable can be written as

$$Y_{M_i} = \frac{M_i - M_i^-}{M_i^+ - M_i^-} \tag{2c}$$

It should be noted here that the interdiffusion coefficients can be estimated by using any of the modified profiles to get the same values. The modified relation can be expressed as

$$\tilde{D}\left(Y_{M_i}^*\right) = \frac{V_M^*}{2t\left(\frac{dY_{M_i}^*}{dx}\right)}\left[\left(1 - Y_{M_i}^*\right)\int_{x^{-\infty}}^{x^*} \frac{Y_{M_i}}{V_M} dx + Y_{M_i}^* \int_{x^*}^{x^{+\infty}} \frac{\left(1 - Y_{M_i}\right)}{V_M} dx\right] \tag{3a}$$

For the intrinsic diffusion coefficients, the ratio of $\frac{D_A}{D_B}$ can be determined using the modified composition profile of A using the modified relation as

$$\frac{D_A}{D_B} = \frac{\overline{V}_A}{\overline{V}_B}\frac{\left[M_A^+ \int_{x^{-\infty}}^{x_K} \frac{Y_{M_A}}{V_M} dx - M_A^- \int_{x_K}^{x^{+\infty}} \frac{1 - Y_{M_A}}{V_M} dx\right]}{\left[-M_B^+ \int_{x^{-\infty}}^{x_K} \frac{Y_{M_A}}{V_M} dx + M_B^- \int_{x_K}^{x^{+\infty}} \frac{1 - Y_{M_A}}{V_M} dx\right]} \tag{3b}$$

If one uses the modified composition profile of B *i.e.* $Y_{M_B}$, it will estimate $\frac{D_B}{D_A}$. Following the absolute values of the intrinsic diffusion coefficients could be estimated using

$$\tilde{D} = C_{M_A}\overline{V}_A D_B + C_{M_B}\overline{V}_B D_A \tag{3c}$$

where $C_{M_i} = M_i / V_M$ is the concentration estimated using the modified composition at the Kirkendall marker plane.

From the relations above, it can be seen that the determination of the molar volume variation is the prerequisite for the estimation of the diffusion parameters. Lattice parameter data are not available in these ternary systems. However, the lattice parameter data available in literature [13-15] indicates that the molar volume varies almost following the Vegard's law in the binary Cu-Sn, Cu-Ga and Cu-Si solid solutions. Therefore, one can estimate the approximate values of the molar volumes at different compositions using

$$V_M = \sum_{i=1}^{n} N_i V_M^i \qquad (4)$$

where $n$ is the number of component and $V_M^i$ is the molar volume of the pure component $i$.

In this study two diffusion couples were prepared. In the Cu-Sn-Ga system, Cu(8Sn) was coupled with Cu(8Ga). In the Cu-Sn-Si system, Cu(8Sn) was coupled with Cu(8Si). Based on the estimated molar volumes of the pure components, $V_M^{Cu} = 7.11$, $V_M^{Sn} = 16.29$, $V_M^{Ga} = 11.80$, $V_M^{Si} = 12.06$ cm$^3$/mol, we estimate the approximated molar volumes of the alloys used for making the diffusion couples as $V_M^{Cu(8Sn)} = 7.84$, $V_M^{Cu(8Ga)} = 7.49$ and $V_M^{Cu(8Si)} = 7.51$ cm$^3$/mol. Since the molar volume variation from one end to the other end of the diffusion couples Cu(8Sn)/Cu(8Ga) and Cu(8Sn)/Cu(8Si) is less than 5%, we consider a constant molar volume variation in our study. Based on our previous analysis on the effect of molar volume on estimated diffusion parameters, we know that this difference gives negligible difference in the estimated diffusion parameters [16]. Therefore, by neglecting molar volume variation (also means $V_M \approx \overline{V}_A \approx \overline{V}_B$), equations 3 can be rewritten as

$$\tilde{D}\left(Y_{M_i}^*\right) = \frac{1}{2t\left(\frac{dY_{M_i}^*}{dx}\right)}\left[\left(1 - Y_{M_i}^*\right)\int_{x^{-\infty}}^{x^*} Y_{M_i} dx + Y_{M_i}^* \int_{x^*}^{x^{+\infty}} \left(1 - Y_{M_i}\right) dx\right] \qquad (4a)$$

$$\frac{D_A}{D_B} = \frac{\left[ M_A^+ \int_{x^{-\infty}}^{x_K} Y_{M_A} dx - M_A^- \int_{x_K}^{x^{+\infty}} \left(1 - Y_{M_A}\right) dx \right]}{\left[ -M_B^+ \int_{x^{-\infty}}^{x_K} Y_{M_A} dx + M_B^- \int_{x_K}^{x^{+\infty}} \left(1 - Y_{M_A}\right) dx \right]} \quad (4b)$$

$$\tilde{D} = N_{M_A} D_B + N_{M_B} D_A \quad (4c)$$

The modified and normalized composition profiles are shown in Figure 2. Following, the estimated interdiffusion coefficients are shown in Figure 3, which are higher in Cu(Sn,Ga) compared to Cu(Sn,Si) over the entire composition range. Although, it is not possible to compare one to one, this result is not surprising, since the interdiffusion coefficients in binary Cu(Ga) is found to be higher than in Cu(Si) [17]. Furthermore, the minimum value is found more or less at the similar composition of 4–4.5 at.% of Ga or Si. Incidentally the Kirkendall markers are found almost at the same composition of ~ 4.2 at.% Ga or Si in both the systems. The ratio of the intrinsic diffusivities at these compositions are estimated as $\frac{D_{Ga}}{D_{Sn}} = 1.3 \pm 0.3$ and $\frac{D_{Si}}{D_{Sn}} = 0.34 \pm 0.06$. Based on the equations 4b and 4c, the intrinsic diffusion coefficients are determined to be $D_{Ga} = 2.29 \times 10^{-14} m^2/s$ and $D_{Sn} = 1.76 \times 10^{-14} m^2/s$ for Cu(Sn,Ga) and $D_{Si} = 3.73 \times 10^{-15} m^2/s$ and $D_{Sn} = 1.09 \times 10^{-14} m^2/s$ for Cu(Sn,Si) system at the Kirkendall plane. Therefore, at the composition of Cu(3.8Sn, 4.2Ga), Ga diffuses with higher rate than Sn, whereas, at the composition of Cu(3.8Sn, 4.2Si), Sn diffuses with higher rate than Si. It should be noted here that the error in estimation of the diffusion coefficients by this method is similar to the binary system. For example, when the ratio falls within the range of 0.1 – 10, the error in estimation is acceptable and might be reasonably higher outside this range [18].

In this study, the procedure for estimating the intrinsic diffusion coefficients utilizing the pseudo-binary approach is formulated. This is explained with the help of the experimental studies in Cu-Sn-Ga and Cu-Sn-Si ternary systems. This approach can also be adopted in higher component systems following the same line of treatment. This approach gives an opportunity to estimate the intrinsic diffusion coefficients in multi-component systems, which was not possible previously. In binary systems, tracer diffusion studies are coupled with interdiffusion studies for insights into the atomic mechanism of diffusion and the role of the thermodynamic driving forces on diffusing components. Now this will be possible to do in multi-component systems as well.

**Acknowledgements:**

Aloke Paul would like to acknowledge the financial support from SERB, India.

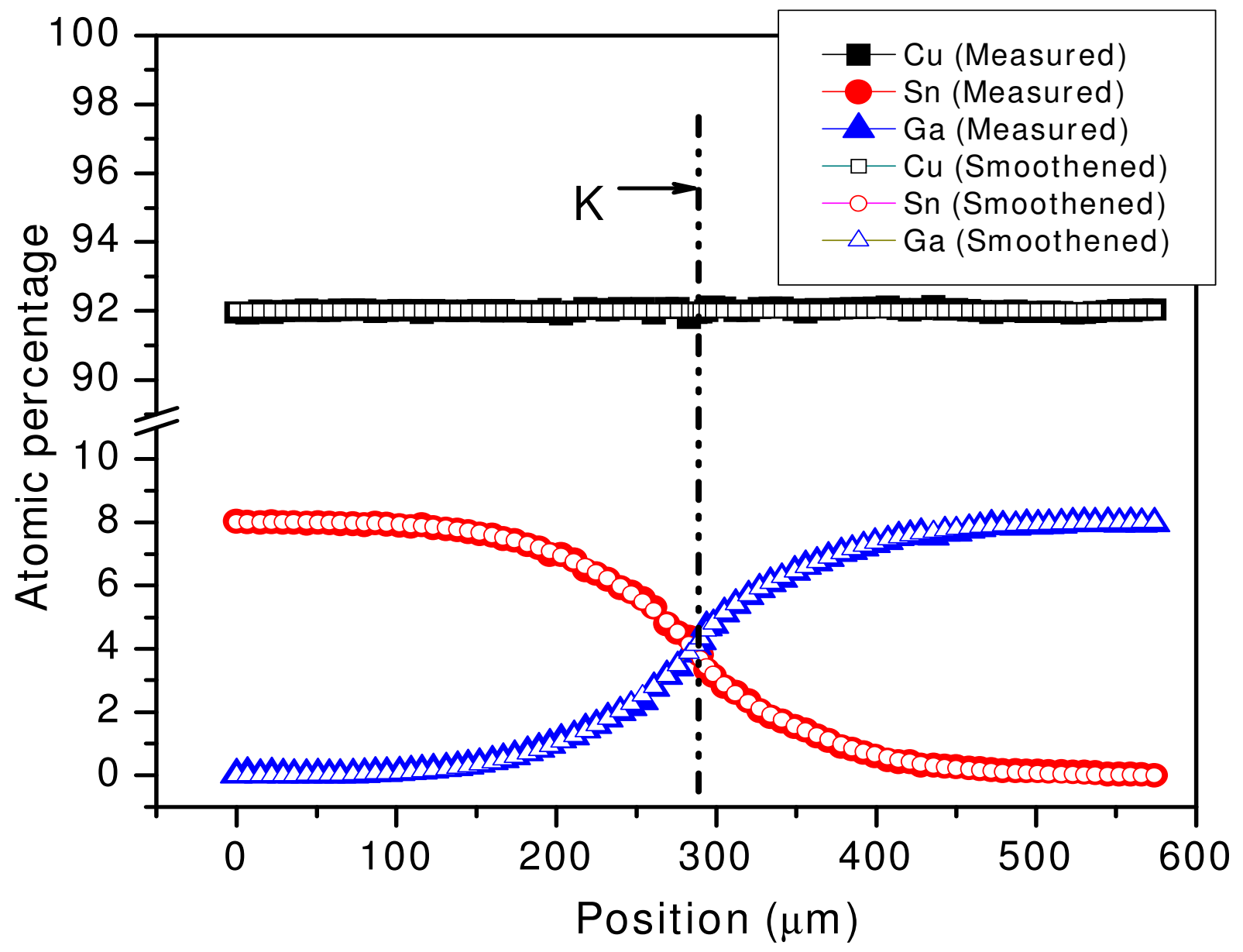

(a)

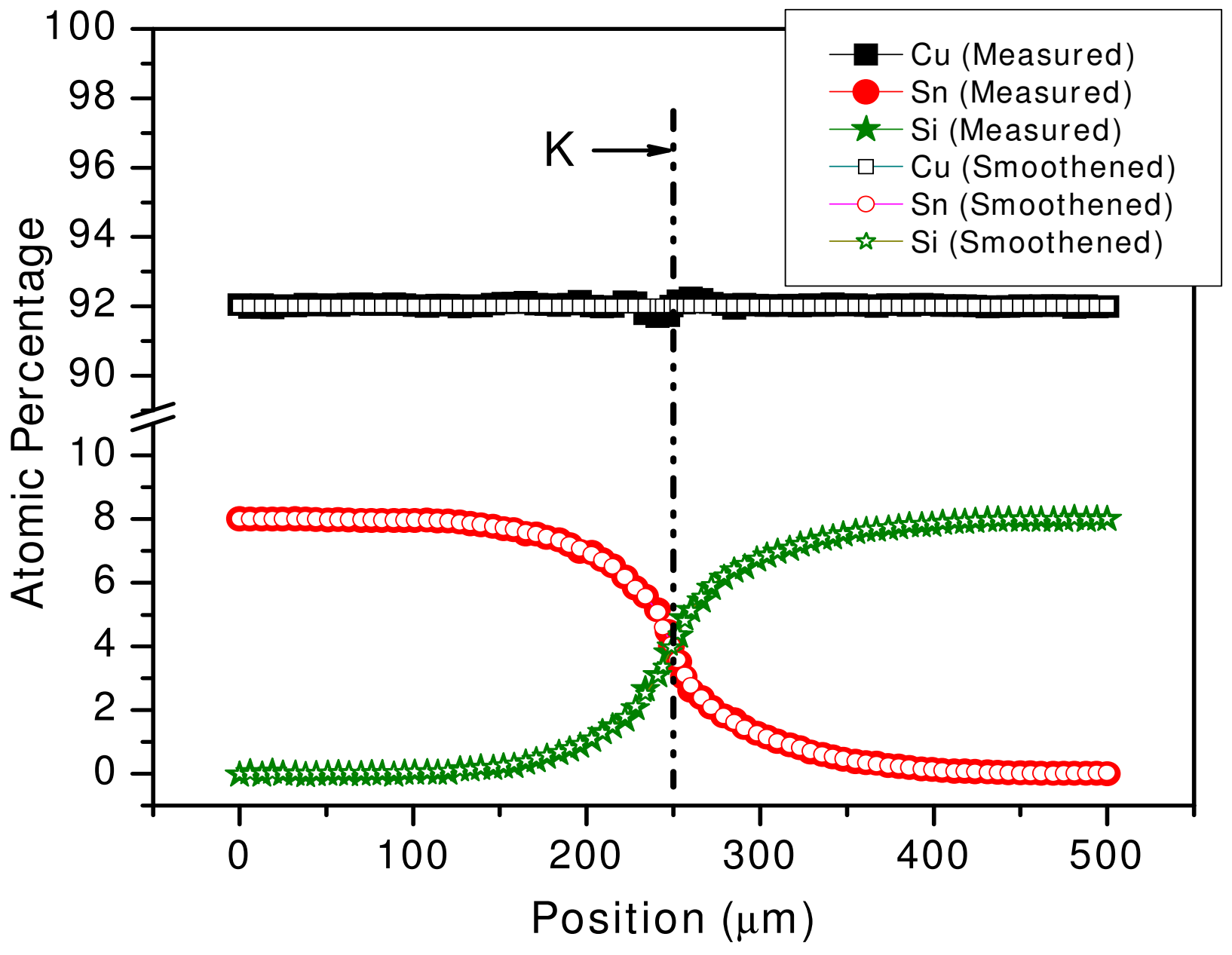

(b)

**Figure 1** Composition profiles developed in (a) Cu(8 at.% Sn)/Cu(8 at.% Ga) and (b) Cu(8 at.% Sn)/Cu(8 at.% Si) diffusion couples after annealing at 700ºC for 25 h. (‘K’ represents Kirkendall plane)

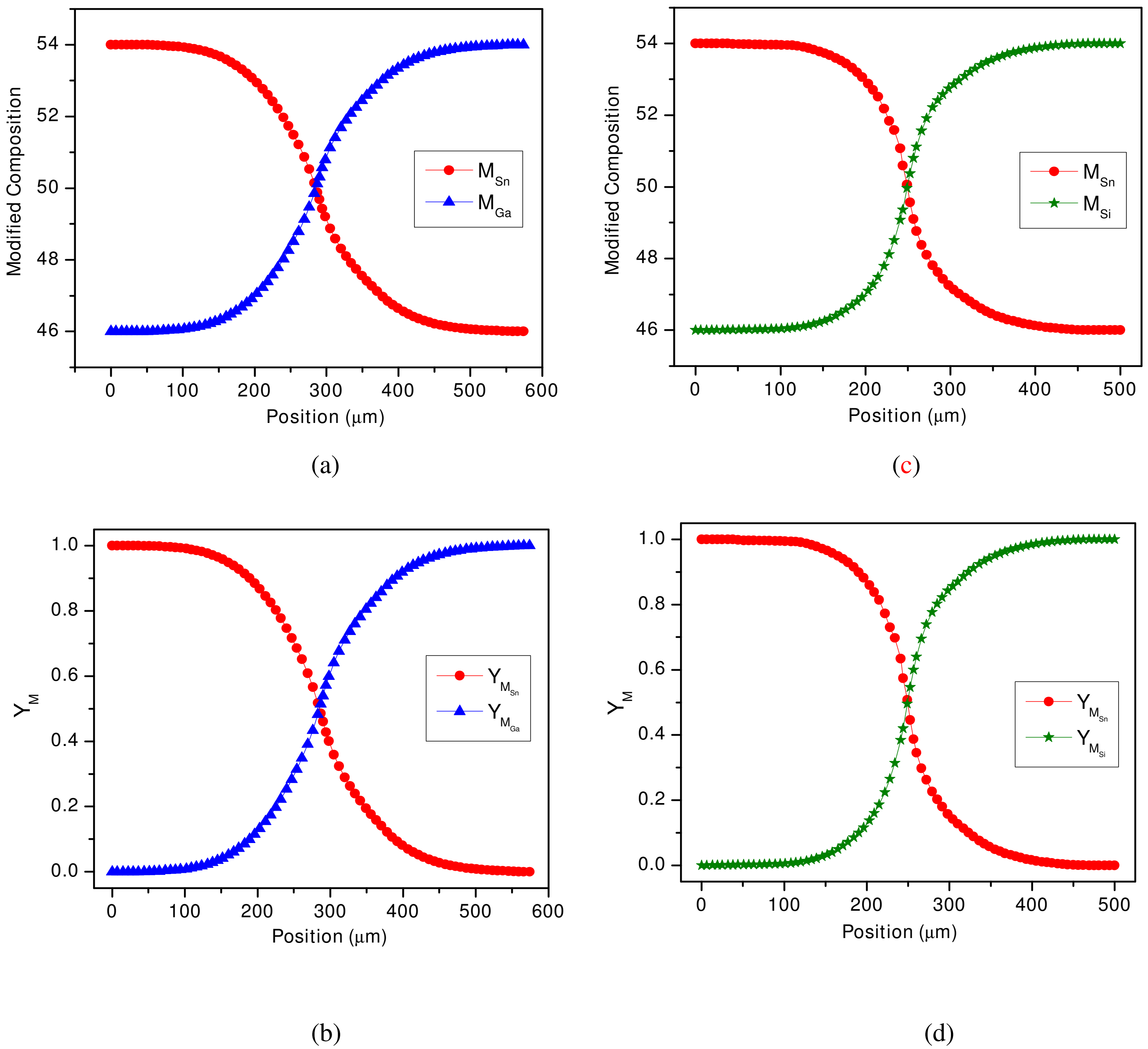

**Figure 2** (a) Modified composition profile and (b) normalized composition profile in Cu(8Sn)/Cu(8Ga) diffusion couple. (c) Modified composition profile and (d) normalized composition profile in Cu(8Sn)/Cu(8Si) diffusion couple.

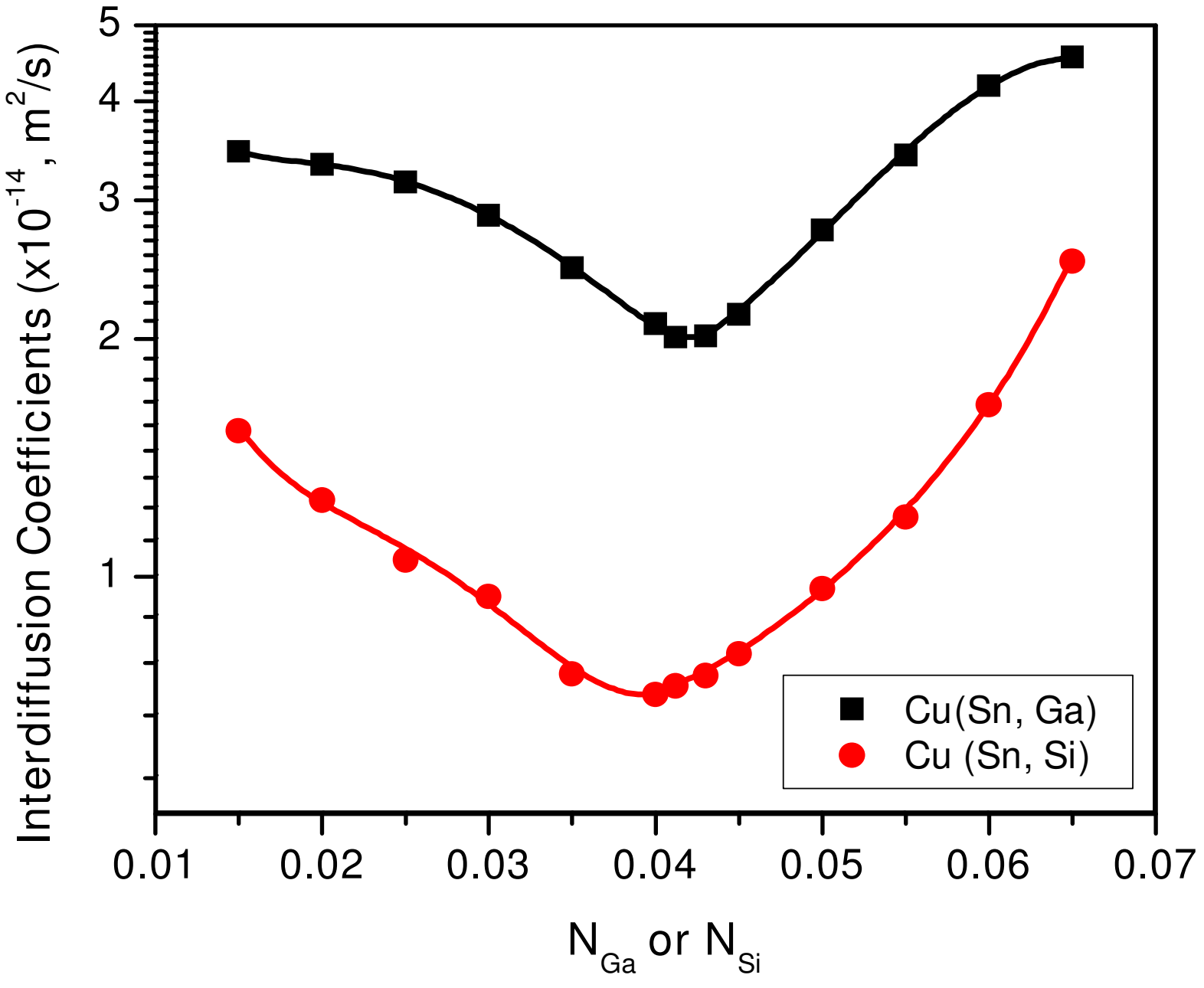

**Figure 3** Variation of estimated interdiffusion coefficients of Sn/Ga and Sn/Si as solutes in a solid Cu solvent with respect to the mole fractions of Ga and Si.